\documentclass{aa}  
\usepackage{diagbox}
\usepackage{graphicx}
\usepackage{xcolor}
\usepackage{subcaption}
\usepackage[fleqn]{amsmath}
\usepackage[utf8]{inputenc}
\usepackage{epsfig,amssymb}
\usepackage[varg]{txfonts}
\usepackage{multirow}
\usepackage{microtype}
\definecolor{mygreen}{rgb}{0.0, 0.6, 0.}
\usepackage[breaklinks,  citecolor=blue, linkcolor=mygreen, urlcolor=purple, colorlinks=true, debug, baseurl=' ']{hyperref}
\usepackage{gensymb}
\usepackage{longtable}

\newcommand{\orcidlink}[1]{\protect\href{https://orcid.org/#1}{\protect\includegraphics[width=8pt]{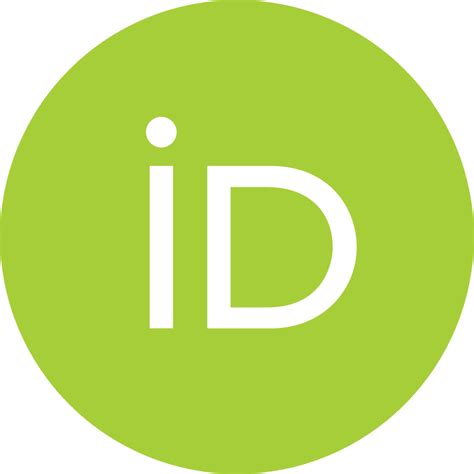}}}

\begin{document}

   \title{3D interstellar medium structure challenges the Serkowski relation}

   \author{N.~Mandarakas
      \inst{1,2}\fnmsep\thanks{nmandarakas@physics.uoc.gr}\orcidlink{0000-0002-2567-2132}
          \and
          K.~Tassis\inst{1,2}\orcidlink{0000-0002-8831-2038}
          \and
          R.~Skalidis\inst{3}\thanks{Hubble Fellow}\orcidlink{0000-0003-2337-0277}
          }

   \institute{Department of Physics, University of Crete, Vasilika Bouton, 70013 Heraklion, Greece
\and
Institute of Astrophysics, Foundation for Research and Technology – Hellas, 100 Nikolaou Plastira, Vassilika Vouton, 70013
Heraklion, Greece
\and
TAPIR, Mailcode 350-17, California Institute of Technology, Pasadena, CA 91125, USA
}

   \date{Received 16 September 2024 / Accepted 10 April 2025}

  \abstract
   {The Serkowski relation is the cornerstone of studies of starlight polarizations  as a function of wavelength. Although empirical, its extensive use since its inception in 1975 to describe polarization induced by interstellar dust has elevated the relation to the status of an indisputable “law”: this is the benchmark against which models of interstellar dust grains are validated.}
   {We revisit the effects of the 3D structure of the interstellar medium (ISM) on the wavelength dependence of interstellar polarization.}
   {We use analytical models to show how the wavelength dependence of both the polarization fraction and direction is affected by the presence of multiple clouds along the line of sight (LOS). We account for recent developments in dust distribution modeling and we utilize an expanded archive of stellar polarization measurements to establish the effect of multiple clouds along the LOS. We highlight concrete examples of stars whose polarization profiles are severely affected by LOS variations in the dust grain and magnetic field properties, and we provide a recipe to accurately fit multiple cloud Serkowski models to such cases.}
   {We present, for the first time, compelling  observational evidence that the 3D structure of the magnetized ISM often results in the violation of the Serkowski relation. We show that 3D effects impact interstellar cloud parameters derived from Serkowski fits. In particular, the dust size distribution in single-cloud sightlines may differ from analyses that ignore 3D effects, with important implications for dust modeling in the Galaxy.}
   {Our results suggest that multiwavelength stellar polarization measurements offer an independent probe of the LOS variations in the magnetic field orientation, and thus constitute a potentially valuable new tool for the 3D cartography of the ISM. Finally, we caution that, unless 3D effects are explicitly accounted for, a poor fit to the Serkowski relation does not, by itself, constitute conclusive evidence that a star is intrinsically polarized.}


   \maketitle

\section{Introduction}
Dust is ubiquitous in the interstellar medium (ISM) of our Galaxy and plays a significant role in many astrophysical processes. The thermal (gray-body) emission of interstellar dust spans many frequencies (100 -- 1000 GHz), 
making dust a powerful tracer of the morphology of ISM structures as well as a diagnostic of their physical properties, such as temperature \citep{hildebrand_1983, andre_2010.herschel}. At the same time, dust emission poses a challenge for cosmological experiments because it veils the cosmic microwave background signal over a wide range of frequencies \citep{planck_2016.component.separation}. Due to the multifaceted role of interstellar dust in astrophysics and cosmology, a detailed understanding of its nature is imperative. 

The interaction of aspherical dust grains with the magnetic field permeating the ISM leads to their alignment, resulting in the polarization of light, which can be observed from ultraviolet to submillimeter wavelengths \citep{andersson_2015.alignment.review}. Short-wavelength (UV, optical, near-IR) polarization is induced by dichroic extinction of the starlight (which usually starts out unpolarized) by aligned dust grains. The same grains emit long-wavelength (far-IR and submillimeter) polarized thermal radiation \citep{Whittet2022,hensley_draine_2023.pahs.astrodust}. Dust polarization provides one of the most reliable and widely employed methods of tracing the properties of the ISM magnetic fields \citep[e.g.,][]{skalidis_tassis2021, skalidis_2021, skalidis_2022, pattle_2023.review, pelgrims_2023, Pelgrims2024}.

Dust-induced stellar polarization follows a complex empirical relation with wavelength, known as the “Serkowski relation” \citep{Serkowski1975}. It expresses the polarization fraction, $P$, as a function of the wavelength, $\lambda$, through the formula
\begin{equation}\label{eq:Serkowski}
    P(\lambda) = P_{max} \cdot \exp \left(-K \cdot \ln^2{\frac{\lambda_{max}}{\lambda}} \right),
\end{equation}
with the use of three free parameters: $P_{max}$, the maximum polarization fraction observed at $\lambda_{max}$, and $K$, which quantifies the spread of the profile. The polarization angle (also referred to as electric vector position angle, EVPA, denoted here as $\theta$) is implicitly assumed to be constant with wavelength.

Since its discovery by \cite{Serkowski1975}, this relation has been extensively studied and confirmed by a number of authors \citep{{Wilking1980,Whittet1992,Whittet2001,Bagnulo2017,Cikota2018}}. It is so well established that it is often dubbed the “Serkowski law.” 
It is even a standard practice in the literature to use it to extrapolate the polarization fraction in one particular wavelength from measurements made at another wavelength \citep{bijas_2024.serkowski.extrapolation, Neha2024, ginski_2024}. Most importantly, the Serkowski relation with its complicated shape provides stringent constraints for dust grain models \citep{guillet_2018.constraining.dust.model.polarization,hensley_2021.observational.dust.constraints.polarization,draine_2024}. 

The Serkowski relation, by construction, describes the polarization induced by a single cloud.
For this reason, in the bibliography, whenever the formula is applied, it is implicitly assumed (even if not stated directly) that the observed (integrated) $P(\lambda)$ signal is induced by a single dominant polarization screen (cloud) along the LOS. Reality, however, is more complicated. \textsc{Gaia} has revealed the third dimension of the sky \citep[depth,][]{gaia_dr2_2018}, and since then it has become clear that in the majority of the LOSs there is more than one cloud with a comparable contribution in the total dust column \citep[e.g.,][]{green_2019.3D.extinction.map,Edenhofer2024}. The variations in the magnetic field along the LOS are imprinted on the individual stellar polarization measurements, allowing for the tomographic mapping of the ISM magnetic field \citep{Pelgrims2024}.

The impact of multiple polarizing clouds along the LOS on stellar polarization has been discussed in the literature, though not extensively. Most of the focus has been on the effect on the EVPA, with less attention given to the polarization fraction spectral profile.
During the early stages of interstellar polarization (ISP) studies, when the notion of a wavelength-independent EVPA emerged from observational evidence, \citet{Treanor1963} briefly mentioned the possibility of a nonconstant EVPA in scenarios involving multiple clouds with different magnetic field alignment and/or different grain sizes.
\citet{Coyne1966} attributed the observed variability in EVPA with wavelength among their targets to the potential presence of multiple clouds along the LOS.
\cite{Martin1974} was the first to construct mathematical models to describe such scenarios, involving multiple clouds with differing properties along the LOS.
Some authors focused on the impact of multiple dust components in the LOS on the parameters derived from Eq.~\ref{eq:Serkowski}.
For example, \cite{CodinaMagalhaes1976} explored the behavior of the $K$ parameter in a medium with a changing grain alignment, while
\cite{Clarke1984} investigated how the presence of two clouds along the LOS would influence the parameters $K$ and $\lambda_{max}$, finding no firm relationship between their models and contemporary observations.

While modern researchers occasionally discuss scenarios of multiple clouds,
most of the focus remains 
on the EVPA \citep[e.g.,][]{McMillanTapia,Messinger1997,Clarke2010,Whittet2015,Bagnulo2017,Whittet2022}. Little attention is given to the broader consequences of such scenarios, in either the $P-\lambda$ profile\footnote{\cite{Patat2010} briefly hints about this in Appendix B.} or the derived dust parameters. 

In this article, we revisit the effects of the 3D structure of the ISM on the wavelength dependence of ISP, accounting for recent developments in dust distribution modeling and utilizing an expanded archive of polarimetric measurements.
We use analytical models to show how both the $P-\lambda$ and $\theta-\lambda$ profiles are expected to behave in scenarios with multiple clouds along the LOS.
For the first time, we present compelling observational evidence to demonstrate these effects and disentangle the polarizing components within the LOS. 
Ultimately,
we demonstrate that the LOS variations in dust and magnetic field properties affect the Serkowski relation in two ways: (1)
they challenge the universality of the relation, because Eq.~\ref{eq:Serkowski} can be a poor fit to the data when 3D effects are significant; and (2) even when Eq.~\ref{eq:Serkowski} fits the data well, the best-fit parameters may not be representative of dust physics. Conversely, deviations of stellar polarization spectral profiles could potentially be utilized as a diagnostic of the 3D complexity of the magnetized ISM.  

\section{Theoretical expectations}
In the case of multiple dust clouds along the LOS, initially unpolarized star light is affected sequentially by all of them, until it reaches the observer.
In the limit of low polarization, the normalized Stokes parameters, $q=Q/I$, $u=U/I$, of the transversing light, are additive \citep[e.g.][]{Patat2010}.
Therefore, the observed $q,u$ at a given wavelength for an intrinsically unpolarized source lying behind $N_C$ number of clouds will be
\begin{equation}\label{eq:add_q}
    q_{obs} = \sum^{N_C}_{i=1} q_i
,\end{equation}
and similarly for $u$, with $q_i$ being the $q$ induced by the $i-th$ dust cloud in the LOS.

In order to explore this effect, we created analytical models for the cases of one and two dust clouds along the LOS, for different parameters of the clouds.
Given that the Serkowski relation is an intrinsic property of individual ISM clouds, variations in grain sizes would be manifested by variations in $\lambda_{max}$, which is proportional to the average size of aligned grains \citep{draine_2024}. Variations in the plane-of-sky (POS) morphology of the magnetic field are traceable through the polarization angle, $\theta$. 
Variations in $\lambda_{max}$ and in $\theta$ between two or more clouds along the LOS significantly affect the cumulative (observed) polarization signal.
For our analytical models, we used Eq.\ref{eq:Serkowski} to calculate $P_i(\lambda)$ for $0.1\mu m \leq \lambda\leq1.2\mu m$, selecting typical values for $P_{max}$ and $\lambda_{max}$ and fixing $K=1.68\cdot \lambda_{max}$ \citep{Wilking1980}.
We also assigned a $\theta_i$ value (constant with wavelength) representing the average POS magnetic field direction within the cloud. Thus, for each cloud, $i$, we have $q_i(\lambda)=P_i(\lambda) \cdot \cos{2\theta_i}$ and $u_i(\lambda)=P_i(\lambda) \cdot \sin{2\theta_i}$.
For the combined model, we calculated $q_{obs}(\lambda)$, $u_{obs}(\lambda)$ from Eq.~\ref{eq:add_q} and finally $\hat{P}(\lambda)$, $\hat{\theta}(\lambda)$ as
\begin{subequations}\label{eq:pevpafromqu}
    \begin{equation}\label{eq:pfromqu}
        \hat{P}(\lambda) = \sqrt{q_{obs}^2(\lambda) + u_{obs}^2(\lambda)}
    \end{equation}
    \begin{equation}\label{eq:evpafromqu}
        \hat{\theta}(\lambda) = \frac{1}{2} \cdot \tan^{-1} \left(\frac{u_{obs}(\lambda)}{q_{obs}(\lambda)}\right),
    \end{equation}
\end{subequations}

where we used the two-argument arctan function.

\begin{figure*}[!htb]
    \centering
    \includegraphics[width=0.9\linewidth]{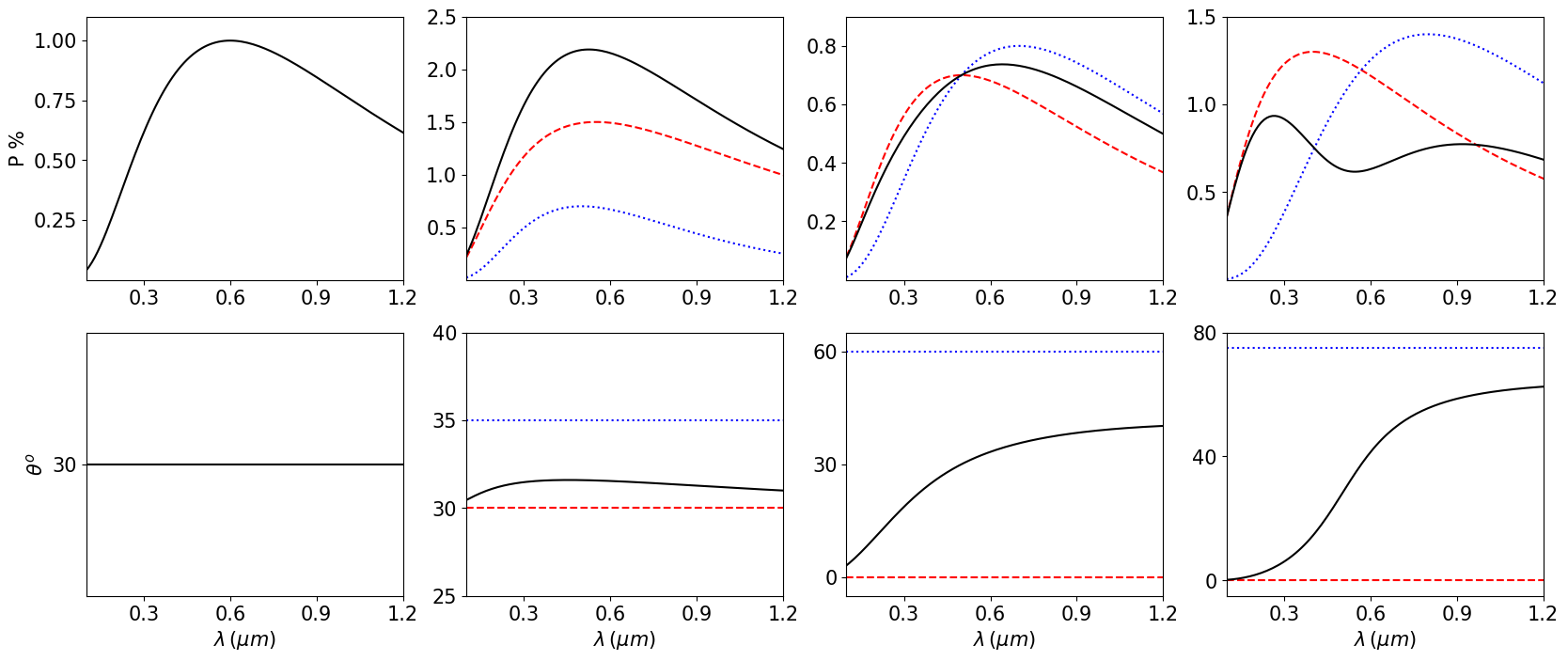}
    \caption{Demonstration of the effect of 3D structure in the Serkowski relation, using analytical models for $P-\lambda$ and $\theta - \lambda$ curves. In each case, the solid black curve corresponds to the combined model and the dashed red and dotted blue lines correspond to individual clouds. Left column: Simple case of one cloud (Eq.~\ref{eq:Serkowski}). Middle left column: Example with two clouds in the LOS, in which both the combined $P-\lambda$, and $\theta-\lambda$ seem to follow the Serkowski relation.
    Middle right column: Example with two clouds in the LOS, in which the combined $P-\lambda$ seems to follow the Serkowski relation, but $\theta-\lambda$ not. Right column: Example with two clouds in the LOS in which both $P-\lambda$ and $\theta-\lambda$ profiles deviate from Serkowski expectations. The parameters for each model are: Left: $P_{max}=1\%$, $\lambda_{max}=0.6\mu m$, $\theta=30^o$. Middle left: $P_{max_1}=1.5\%$, $\lambda_{max_1}=0.55\mu m$, $\theta_1=30^o$, $P_{max_2}=0.7\%$, $\lambda_{max_2}=0.5\mu m$, $\theta_2=35^o$. Middle right: $P_{max_1}=0.7\%$, $\lambda_{max_1}=0.5\mu m$, $\theta_1=0^o$, $P_{max_2}=0.8\%$, $\lambda_{max_2}=0.7\mu m$, $\theta_2=60^o$. Right: $P_{max_1}=1.3\%$, $\lambda_{max_1}=0.4\mu m$, $\theta_1=0^o$, $P_{max_2}=1.4\%$, $\lambda_{max_2}=0.8\mu m$, $\theta_2=75^o$. For all clouds, we have used $K=1.68 \cdot \lambda_{max}$ \citep{Wilking1980}}.
\label{fig:toymodels}
\end{figure*}

In Fig.~\ref{fig:toymodels}, we present some examples of $P-\lambda$ and $\theta - \lambda$ profiles resulting from the presence of one or two clouds in the LOS, explored by our analytical models.
The simple case of a single dust cloud (Serkowski relation - Eq.\ref{eq:Serkowski}) is shown in the left panel of Fig.~\ref{fig:toymodels} for reference.
Three cases with two clouds in the LOS are shown in the middle left, middle right, and right panels of Fig.~\ref{fig:toymodels}.
Specifically,the middle left panel demonstrates an example where both the $P-\lambda$ and $\theta-\lambda$ curves resemble the Serkowski relation, although the profiles are constructed by two components.
The middle right panel demonstrates an example where the $P-\lambda$ curve resembles the Serkowski relation, yet the $\theta-\lambda$ profile deviates from constancy. The right panel demonstrates an example where both the $P-\lambda$ and the $\theta - \lambda$ profiles deviate from the Serkowski curves. 

Assuming a difference of a factor of two in $\lambda_{max}$ between our two hypothetical clouds yields a bimodal $P(\lambda)$, which is significantly different than the shape of a typical Serkowski curve (Fig.~\ref{fig:toymodels} upper right panel). Although the difference appears to be significant, it would be challenging to identify it in practice without the use of spectropolarimetry.
Current polarimetric research on the ISM predominantly employs wide-filter polarimetry, which has limited resolving power. Moreover, studies with multiband polarization measurements (such as the ones used in this work - see Sect.~\ref{subsec:data}) are usually limited to a narrow range of frequencies.

Modest LOS variations in $\lambda_{max}$ can result in even more complications. In our assumed case of two clouds, a $40\%$ difference in $\lambda_{max}$ between the two clouds leads to a profile that is nearly identical to the single cloud Serkowski curve profile (middle right panel in Fig.~\ref{fig:toymodels}), although the cumulative signal by construction consists of two underlying profiles.
In this case, erroneously assuming a single cloud along the LOS would lead to inaccurate constraints for the parameters of the Serkowski relation, and hence for the grain properties.
The imprint of a second cloud in these cases can be noticed by the nonconstant behavior of $\theta$ with wavelength.

In cases in which the $\lambda_{max}$ and $\theta$ parameters are similar between clouds (middle left panel in Fig.~\ref{fig:toymodels}), the resulting profiles of both polarization and EVPA are indistinguishable from the single cloud case within observational uncertainties, given the current accuracy of optical polarimeters.
Significant differences in the EVPAs of two clouds lying along the LOS, such as the ones employed in the middle right and rightmost models of Fig.~\ref{fig:toymodels},
can be revealed through the variation in the EVPA as a function of $\lambda$. However, obtaining a sufficient number of data points that would allow us to probe such variations is challenging\footnote{Spectropolarimetry is one of the few robust methods of accurately probing $P$ or EVPA variations with wavelength \citep[e.g.,][]{Bagnulo2017}, but so far this method has only been applied to a very limited number of stars.}, and for this reason these data are sparse. 

These analytic calculations pose a challenge in the interpretation of $P(\lambda)$ observations: while the Serkowski relation would still hold for individual clouds, the majority of the results extracted from multiband polarization are likely contaminated by the 3D structure of the ISM, leading to inaccurate dust-grain constraints. The upside of this problem is that it opens up new possibilities for studying the magnetized ISM in 3D.

\section{Methods}

\subsection{Polarimetric data selection -- Statistical sample}\label{subsec:data}
A recent agglomeration of multiband optical polarization data \citep{Panopoulou2025}, which is the largest to date, allows us, for the first time, to study in a statistical fashion the impact of the 3D ISM structure of the Galaxy on the Serkowski relation, and the consequences for magnetic field and dust physics studies.
In order to fit multiple-cloud models to polarization data, it is optimal to use polarization measurements across a broad optical wavelength range. However, for most stars in the literature,  measurements exist only in a single band or a few bands.
Even so, it is possible to evaluate the validity of the Serkowski relation in a statistical manner, by studying an ensemble of sources. We designed such an experiment to test whether stars in single-cloud LOSs are better fit by the Serkowski relation than stars in multiple-cloud LOSs.

To acquire archival optical polarization measurements for the statistical comparison, we explored Tables 4 and 6 of the compiled polarization catalog of \cite{Panopoulou2025}, which provide polarimetry and distances from \textsc{Gaia} EDR3 \citep{BailerJones2021}, respectively, for $\sim 42,000$ sources.
We excluded stars with distances greater than 1.25 kpc, as this is the limit of our dust map (see Sect.~\ref{subsec:extinction}).
As a distance, we took the average of the values (`r\_med\_geo',`r\_med\_photogeo') provided in the catalog.
We excluded measurements in filters with no well-defined effective wavelength ($\lambda_{eff}$); 
therefore, the excluded filters are “0,” “20,” “23,” and “51,” which correspond to “No filter or unclear,” “weighted-mean,” 0.735-0.804 $\mu m$, and 0.5-0.85 $\mu m$, respectively, in the catalog.
For targets with multiple measurements in the same filter, we used the weighted mean and standard deviation of the $q$ measurements, according to
\begin{equation}\label{eq:w.mean}
    q_{\text{mean}} = \frac{\sum(q_i \cdot w_i)}{\sum w_i} \,\, 
\end{equation}
\begin{equation}\label{eq:w.std}
    q_{\text{std}} = \frac{\sum w_i (x_i - \bar{x})^2}{\left(\frac{N-1}{N}\right) \cdot \sum w_i}, 
\end{equation}
with $w_i=1/\sigma_{qi}^2$, and $\sigma_{qi}$ the uncertainty of the measurement in $q$, and similarly for $u$.
We excluded measurements with debiased $P/\sigma_P<2$.
Additionally, we did not consider measurements with a polarization uncertainty of $\sigma_P<0.1\%$ because they are likely underestimated.
After all the cuts, we selected targets that have surviving measurements in at least three bands.
We note that we did not impose any cut for intrinsically polarized stars (although the \cite{Panopoulou2025} catalog provides this information), as these targets may have been flagged as intrinsically polarized simply because they do not follow the Serkowski relation.
However, the final sample does not contain any star that is flagged as intrinsically polarized.
The final sample contains 223 sources with measurements in at least three optical bands.

\begin{figure*}[tbp]
    \centering
    \begin{subfigure}[b]{0.49\linewidth}
        \centering
        \includegraphics[width=\linewidth]{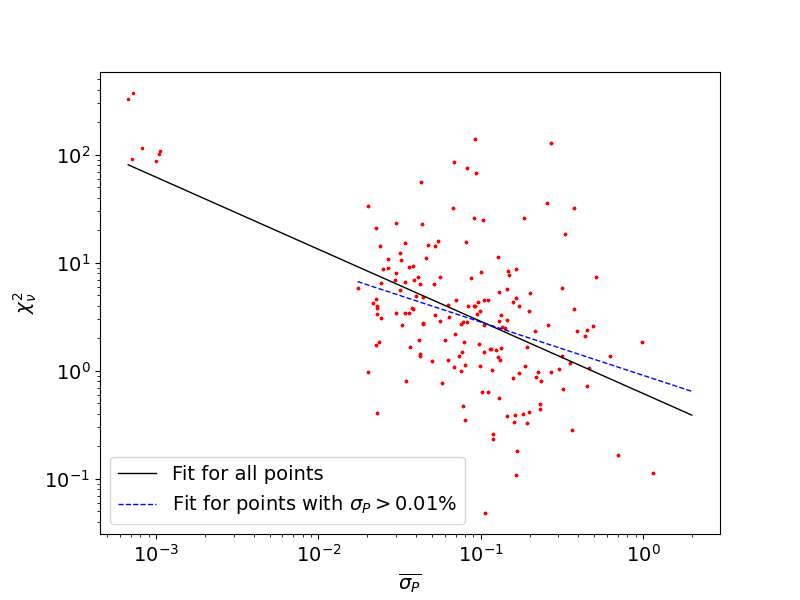}
        \label{fig:chierrall}
    \end{subfigure}
    \hfill
    \begin{subfigure}[b]{0.49\linewidth}
        \centering
        \includegraphics[width=\linewidth]{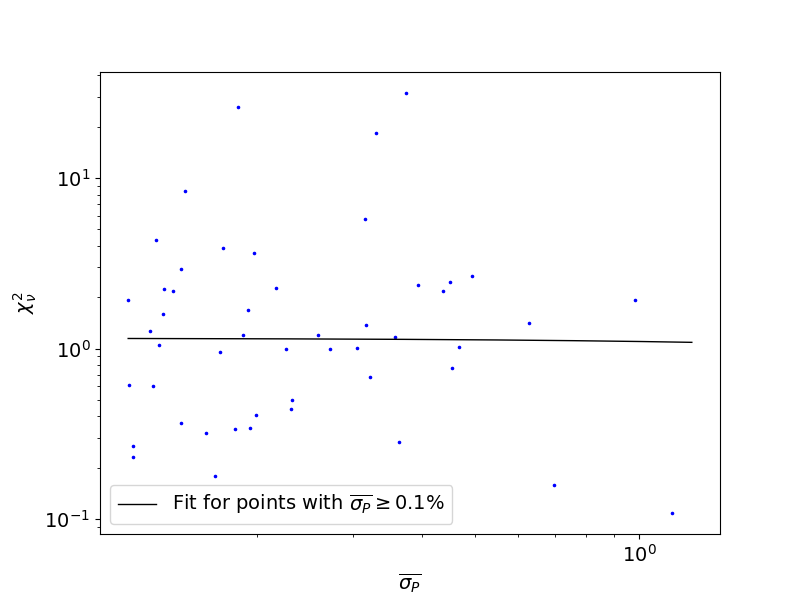}
        \label{fig:chierrgood}
    \end{subfigure}
    \caption{Left: Average reported uncertainties in polarization, $\overline{\sigma_P}$, versus $\chi^2_\nu$ for targets that survive the cuts described in Sect.~\ref{subsec:data}, but without any restriction for the error bars. The black line is a linear fit for all the data in log-log space. The dashed blue line is similar, but excluding the points in the top left of the plot. Right: Similar to left, but with the restriction for the error bars of $\sigma_P\geq0.1\%$.}
    \label{fig:combined}
\end{figure*}

The reasoning for excluding data with $\sigma_P<0.1\%$ is presented below.
Initially, we considered not imposing any restrictions on the level of $\sigma_P$. However, we found that error bars below $\sim0.1\%$ are likely underestimated on average. If the error bars accurately represented the uncertainties in the measurements, the $\chi^2_\nu$ would remain independent of the error bar level when analyzing targets within the same subsample (i.e., those behind the same number of clouds). To test this hypothesis, we examined the targets in the subgroup with $N_C=1$. We chose this particular subsample, as we know not only that the $\chi^2_\nu$ should be flat with respect to the error bars, but also that it should be $\sim 1$ for targets behind a single cloud. We fit the model described in Sect.~\ref{subsec:statistics} and, for each target, plotted the average error bar value across different filters against the corresponding $\chi^2_\nu$. We demonstrate in the left panel of Fig.~\ref{fig:combined} that targets with smaller error bars exhibit higher $\chi^2_\nu$, revealing a clear, nonconstant correlation between the two quantities. $\chi^2_\nu$ seems to approach the expected value of $\sim1$ for $\sim\sigma_P\geq0.1\%$. Based on the above findings, we gradually excluded points with lower reported uncertainties and repeated the fit described in Sect.~\ref{subsec:statistics}, until the expected behaviors were satisfied. We found that the dependence practically disappears and $\chi^2_\nu$ approaches the expected value of $\sim1$ for targets with average $\sigma_P\geq0.1\%$ (Fig.~\ref{fig:combined}, right). Therefore, we discarded all measurements with $\sigma_P<0.1\%$.

\subsection{Extinction}\label{subsec:extinction}
We collected extinction data for our target stars from the \cite{Edenhofer2024} dust map.
We chose this map, as it provides a good angular resolution (14') as well as a parsec-scale distance resolution.
The map is available in two versions, one that extends out to 1.25 kpc from the Sun and one that extends up to 2 kpc, but uses less data for the dust profile reconstruction.
We used the 1.25 kpc version, as it should be more accurate, and we did not want to introduce any additional sources of noise.
Extinction values in the \cite{Edenhofer2024} map are expressed in the units used by \cite{Zhang2023}, the work upon which the dust map was based.
We converted the extinction values to represent the extinction in the $V$ band ($A_V$) based on the conversion table provided by \cite{Zhang2023}.

\begin{figure*}
    \includegraphics[width=1.0\textwidth]{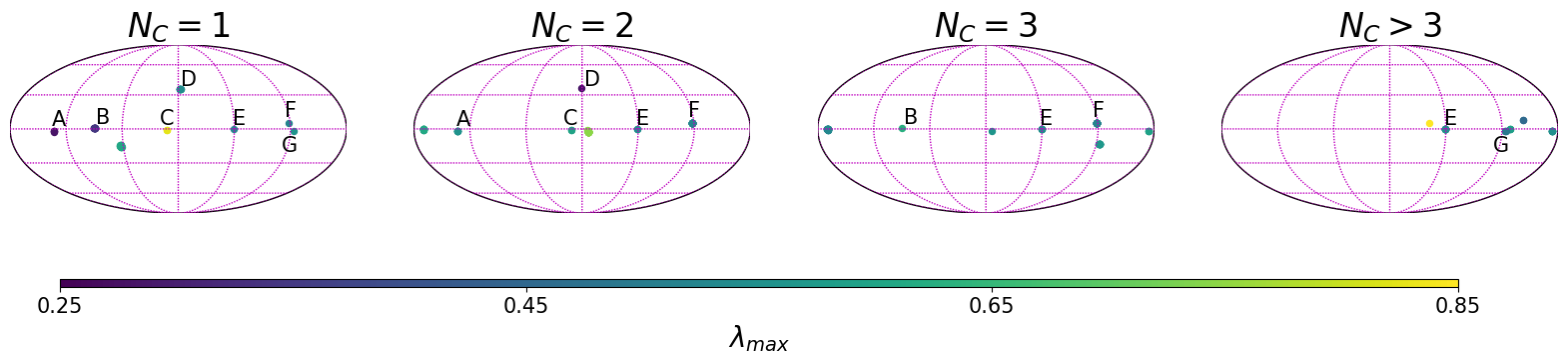}
    \caption{Positions of targets depending on their $N_C$. Maps are in Galactic coordinates, in mollweide projection, centered on (0,0). Parallels are drawn every 30$^o$. Meridians are drawn every 60$^o$. Longitude increases toward the right-hand side. Colors correspond to the average $\lambda_{max}$ of each area. Areas marked with “A” through “G” contain stars that are located behind a single cloud and multiple clouds. See discussion in Sect.~\ref{subsec:parameters}.}
    \label{fig:skypositions}
\end{figure*}

For each of our targets, we examined the “mean” \citep{Edenhofer2024} cumulative and differential $A_V$ versus distance ($D$) profiles by eye, in order to identify the number of significant steps in the profile, which presumably correspond to distinct dust clouds ($N_C$).
For targets for which more than three clouds were identified, or the dust profile was very complex, we set $N_C>3$.
The number of target stars in each subgroup with a different number of identified clouds along the LOS are $\#(N_C=1)=49$, $\#(N_C=2)=74$, $\#(N_C=3)=70$, $\#(N_C>3)=30$.
The targets that correspond to each $N_C$ are distributed over different positions on the sky (Fig.~\ref{fig:skypositions}).

We considered the possibility of using a more objective metric to characterize the number of dust clouds along each LOS, as each step in the dust profiles is typically not equally prominent, and in many cases numerous small steps contribute to the cumulative extinction. Therefore, we explored a method similar to the one used by \cite{Panopoulou2020}, who identified the number of clouds in \textsc{Hi} data by detecting peaks in the differential \textsc{Hi} column density profile and then normalizing each peak by the height of the tallest peak among them. In our case, the normalized number of clouds in each LOS would be $N_C^{norm} = \sum \frac{A_V^i}{A_V^{max}}$, with $A_V^i$ the extinction of a given peak in the LOS, and $A_V^{max}$ the extinction of the highest peak in the differential $A_V$ versus $D$ profile. However, this was not useful in our study for the following reasons. 1) $N_C^{norm}$ is ambiguous. For example, $N_C^{norm}=2$ could signify two equally prominent peaks, or one prominent peak plus two peaks half its height, or even one prominent peak plus three peaks a third of its height, and so on. This lack of clarity reduces its utility. 2) The profiles in the \cite{Edenhofer2024} map are sometimes irregular, with many minor peaks that may not represent real features but instead result from sampling noise. Additionally, peaks can appear very close together, making it difficult to distinguish between separate clouds and variations within the same cloud. Although we considered smoothing the profiles to address these issues, we found that the number and relative height of the resulting peaks were highly sensitive to the parameters chosen for smoothing, introducing further ambiguity. 3) Even when one peak appears higher than another, this does not necessarily correlate with increased polarization. The dust-induced polarization of aligned grains depends largely on factors beyond peak height, such as dust density, grain alignment efficiency, magnetic field coherence, and the inclination of the magnetic field with the LOS.

\begin{figure*}[ht]
    \centering
    \begin{subfigure}[b]{0.24\textwidth}
        \includegraphics[height=5.5cm,width=\textwidth]{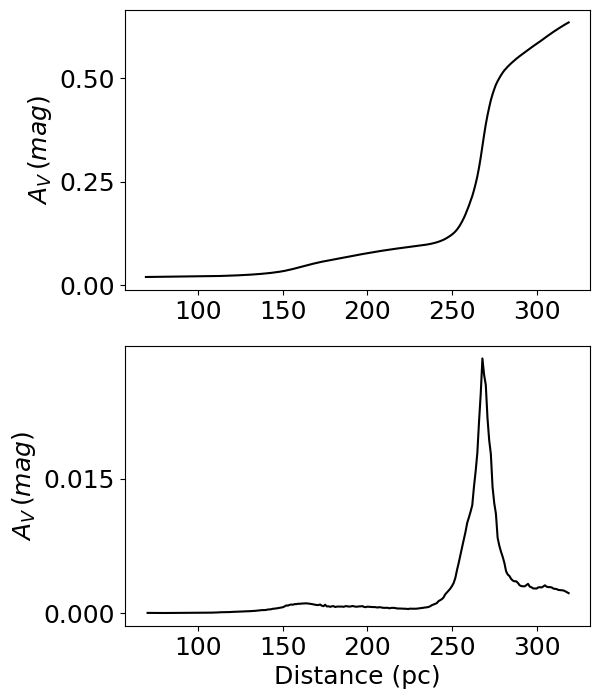}
    \end{subfigure}
    \begin{subfigure}[b]{0.24\textwidth}
        \includegraphics[height=5.5cm,width=\textwidth]{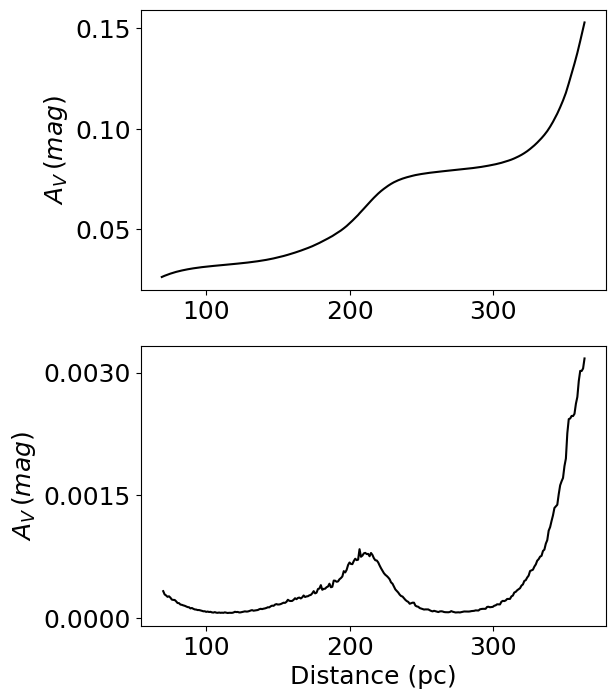}
    \end{subfigure}
    \begin{subfigure}[b]{0.24\textwidth}
        \includegraphics[height=5.5cm,width=\textwidth]{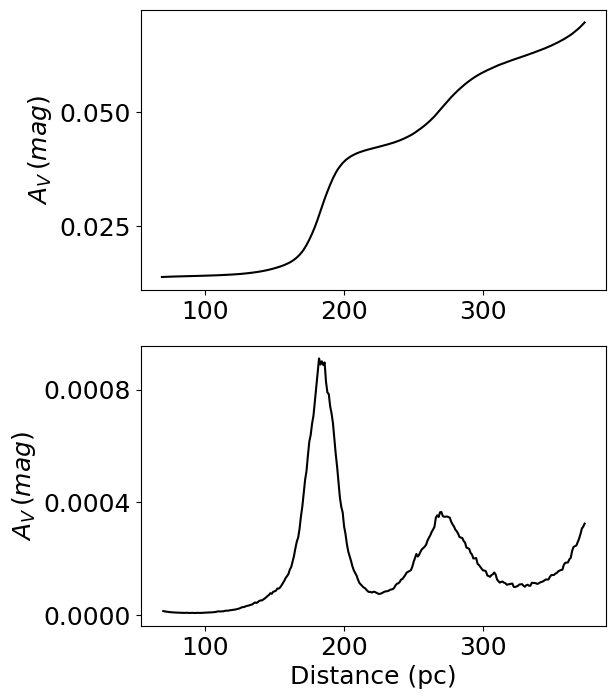}
    \end{subfigure}
    \begin{subfigure}[b]{0.24\textwidth}
        \includegraphics[height=5.5cm,width=\textwidth]{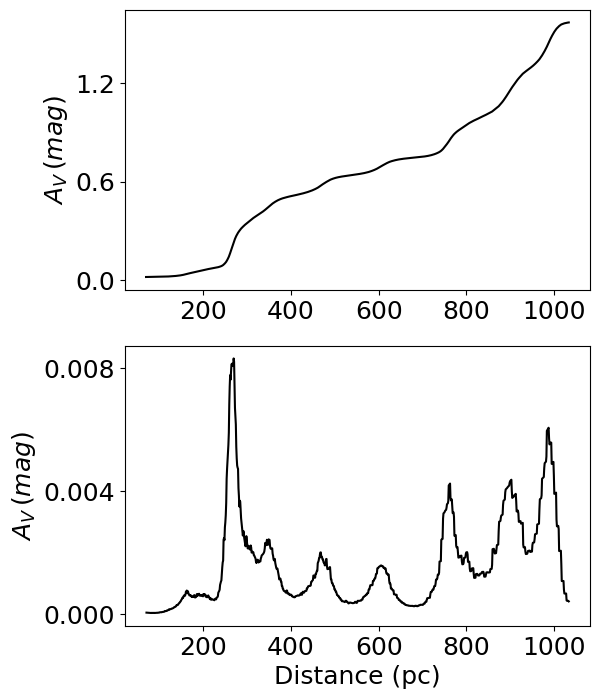}
    \end{subfigure}
    
    \caption{Dust profiles for four example targets belonging in four different subsamples. From left to right: $N_C=1$, $N_C=2$, $N_C=3$, $N_C>3$. Top row: Cumulative $A_V$ vs. $D$. Bottom row: Differential $A_V$ vs. $D$.}
    \label{fig:dust_profiles}
\end{figure*}

In our attempts to devise an objective method of peak counting, we found that the resulting numbers were often ambiguous or lacked physical meaning. Consequently, we opted for a visual approach. Although this method has its uncertainties, it is self-consistent and allows statistical patterns to emerge in a dataset as large as ours. Figure~\ref{fig:dust_profiles} illustrates examples of dust profiles that we assigned to different subsamples.

\subsection{Fitting the Serkowski relation}\label{subsec:fitmethod}
In order to evaluate the performance of the Serkowski relation in our statistical sample,
we used an extremely powerful, if unconventional in Serkowski studies, technique: we fit our data in the Stokes parameters space $q-u$, exploiting its power to simultaneously trace changes in the profiles of both the degree of polarization and EVPA.
Traditionally, the EVPA is not taken into account while fitting the Serkowski formula, and thus part of the information imprinted in the polarization signal is ignored.
Although the variability of the EVPA with wavelength has been hypothesized to be connected to the existence of multiple clouds \citep[e.g.][]{McMillanTapia,Whittet2015}, $q-u$ fitting has never been employed to trace the various components (clouds) contributing to the $P-\lambda$ and $\theta-\lambda$ profiles.

As a first step, we converted Eq.~\ref{eq:Serkowski} to the $q-u$ space.
This step allows us to include $\theta$ in the model, which is ignored in the Serkowski formula.
Therefore, our model becomes
\begin{subequations}\label{eq:quhat}
\begin{equation}\label{eq:q_hat}
    \hat{q} = P_{max} \cdot \exp \left(-K \cdot \ln^2{\frac{\lambda_{max}}{\lambda}} \right) \cdot \cos(2\theta),
\end{equation}
\begin{equation}\label{eq:u_hat}
    \hat{u} = P_{max} \cdot \exp \left(-K \cdot \ln^2{\frac{\lambda_{max}}{\lambda}} \right) \cdot \sin(2\theta).
\end{equation}
\end{subequations}
The quantity we wished to minimize for the fits is
\begin{equation}\label{eq:m}
    m = L_q + L_u + \frac{L_q}{L_u} + \frac{L_u}{L_q}
,\end{equation}
with
\begin{subequations}
\begin{equation}\label{eq:L_q}
    L_q = \frac{1}{N_{\lambda}} \cdot \sum \left( \frac{q_i - \hat{q}_i}{\sigma_{q_i}} \right)^2,
\end{equation}
\begin{equation}\label{eq:L_u}
    L_u = \frac{1}{N_{\lambda}} \cdot \sum \left( \frac{u_i - \hat{u}_i}{\sigma_{u_i}} \right)^2,
\end{equation}
\end{subequations}
where $N_{\lambda}$ is the number of different bands for which we had measurements for each target. $q_i$, $u_i$ is the measurement in a given band, and $\hat{q_i}$, $\hat{u_i}$ our model introduced in Eq.~\ref{eq:q_hat}, \ref{eq:u_hat}. $\sigma_{q_i}$, $\sigma_{u_i}$ are the uncertainties of the measurements in $q$ and $u$, respectively.
The terms $L_q$ and $L_u$ are the reduced $\chi^2$ metric for $q$ and $u$, respectively.
The terms $L_q/L_u$ and $L_u/L_q$ were introduced to ensure a balanced contribution from $L_q$ and $L_u$.
For example, if the objective was to minimize just $L_q + L_u$, the optimal value $V$ could be achieved with  $L_q=V$ and $L_u=0$, $L_u=V$ and $L_q=0$, or any combination in between.
By introducing $L_q/L_u$ and $L_u/L_q$, we penalize the model when it favors solutions where $L_q=0$ or $L_u=0$, and thus promote a more balanced fit.

For the modeling procedure, we used the \textsc{Optuna} package in Python \citep{Akiba2019}.
\textsc{Optuna} is an automatic hyperparameter optimization software framework, originally developed for machine learning hyperparameter-tuning purposes.
However, it can also address general optimization problems through the use of a Bayesian optimization framework.
We found it highly effective in multiparameter optimization with few data points and many parameters, such as the case of the $q-u$ modeling, and more efficient in finding the best solution than traditional algorithms such as Markov chain Monte Carlo (MCMC).
We used the default options of \textsc{Optuna} with n\_trials = 5000 for our fits.
Moreover, we adopted the following priors.
For $P_{max}$, we initially assumed a rather loose Gaussian prior distribution with a mean of $3\%$ and standard deviation of $3\%$, and enforced it to be positive.
For $\lambda_{max}$, we assumed a Gaussian prior distribution with a mean of 0.6 $\mu m$ and standard deviation of 0.2 $\mu m$, and ensured it was always positive.
For $K$, we imposed a Gaussian prior distribution with a mean of $1.68\cdot \lambda_{max}$ and standard deviation of $0.13\cdot \lambda_{max}$ \citep{Wilking1980}, and ensured it was always positive.
For $\theta$, we assumed a Gaussian prior distribution with the mean the value derived from the average $q$ and $u$ as in Eq.~\ref{eq:w.mean} and converted to $\theta$ using Eq.~\ref{eq:evpafromqu},
and a $10^o$ standard deviation.

We evaluated the quality of our fits by measuring the $\chi^2$ per degree of freedom (DoF):
\begin{equation}\label{eq:chi_dof}
    \chi^2_{\nu} \equiv \frac{\chi^2}{DoF} =  \frac{\sum \left( \frac{q_i - \hat{q}_i}{\sigma_{q_i}} \right)^2 + \sum \left( \frac{u_i - \hat{u}_i}{\sigma_{u_i}} \right)^2}{2\cdot N_{\lambda} - \nu},
\end{equation}
where $\nu=4$ is the number of free parameters in the model ($P_{max}$, $\lambda_{max}$, $K$, $\theta$).
We used $2\cdot N_{\lambda}$ as the number of points in Eq.~\ref{eq:chi_dof} as we accounted for the measurements in both $q$ and $u$ to perform the simultaneous fits.
We note that we cannot use $m$ directly (Eq.~\ref{eq:m}) to evaluate the fits, as $m$ is not a normalized metric, and thus cannot be used for comparing fits to each other.

\subsection{Fitting multiple clouds}\label{subsec:multiple_component_fits}
In order to fit multiple clouds (as in the examples of Sect.~\ref{subsec:examples}), 
we again used the $q-u$ fitting technique, following the same logic discussed in Sect.~\ref{subsec:fitmethod}.
The multiple cloud model is derived from Eqs.~\ref{eq:add_q} and \ref{eq:quhat}, by adding up the $q$ and $u$ components of the individual clouds.

When trying to determine the correct parameters for a model with two or more clouds, the number of parameters is increased (four parameters per cloud), resulting in a substantial number of possible combinations and a vast parameter space to explore.
In such situations, identifying the optimal combination of parameters that minimize the cost function can be challenging.
To address this, we employed the following procedure.
Initially, instead of directly using \textsc{Optuna}, we utilized an MCMC algorithm to rapidly explore the parameter space. Given that the two clouds are interchangeable, we imposed the condition $\lambda_{max_1}>\lambda_{max_2}$ to focus the algorithm on a single solution.
Subsequently, we visually examined the corner plots of the parameters generated by the MCMC algorithm to establish reasonable constraints for the \textsc{Optuna} algorithm, thereby significantly narrowing the search space.
Finally, we ran the fit using \textsc{Optuna}, imposing Gaussian priors on the parameters, based on the mean and the spread of the parameters identified by the MCMC algorithm.
We chose \textsc{Optuna} to derive the final set of parameters for the model, as it directly provides the optimal set of parameters by minimizing Eq.~\ref{eq:m}.
The fits were evaluated with $\chi^2_\nu$ (Eq.~\ref{eq:chi_dof}).

\section{Results}

\subsection{Statistical sample}\label{subsec:statistics}

\begin{figure*}
\centering
\begin{subfigure}{0.49\textwidth}
    \includegraphics[width=\textwidth]{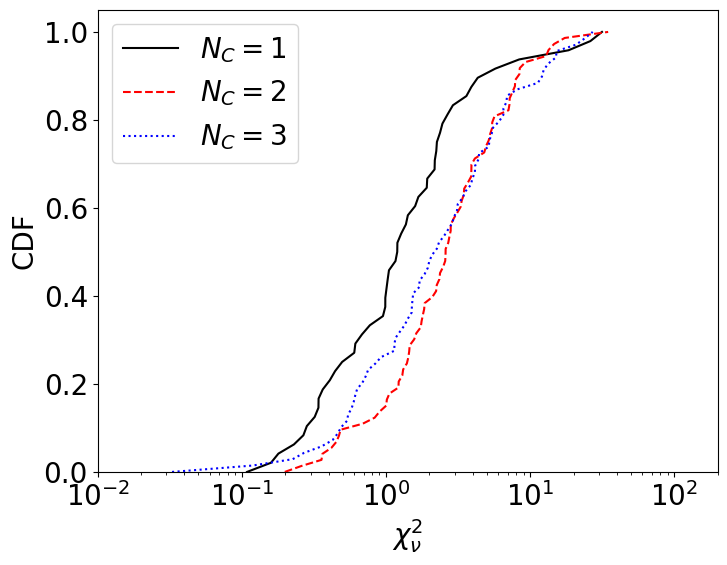}
\end{subfigure}
\hfill
\begin{subfigure}{0.49\textwidth}
    \includegraphics[width=\textwidth]{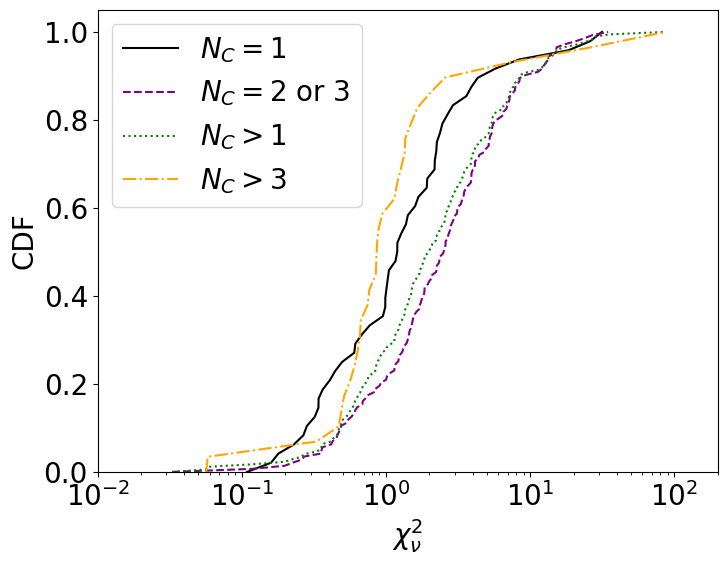}
\end{subfigure}
\caption{CDFs of $\chi^2_{\nu}$ for different subsamples, as is described in each label.}
\label{fig:CDFs}
\end{figure*}

We grouped the sample of sources found in Sect.~\ref{subsec:data} into subsamples depending on the number of clouds ($N_C$) lying in their LOS, as derived from the 3D extinction map. We obtained four main subsamples: targets in LOSs with one dominant cloud, $N_C = 1$; two clouds, $N_C = 2$; three clouds $N_C = 3$; and more than three clouds, $N_C > 3$.
For this sample, we fit the traditional Serkowski curve (Eq.~\ref{eq:Serkowski}) for all targets, regardless of $N_C$, to evaluate the capability of the Serkowski relation to fit polarimetric data for subsamples with different $N_C$.
We evaluated the performance of the Serkowski relation in all of the targets as described in Sect.~\ref{subsec:fitmethod}.
We constructed the cumulative density functions (CDFs) of $\chi^2_{\nu}$ for the different subsamples (Fig.~\ref{fig:CDFs}), and 
we applied the Kolmogorov-Smirnov (KS) test
to identify any statistical discrepancies between the $\chi^2_\nu$ distributions of the $N_C = 1$ subset and the other three subsets.

We found that the  $\chi^2_\nu$ distribution of the $N_C=1$ subset is significantly different from those of subsets with $N_C=2$ or $N_C=3$.
In contrast, the $\chi^2_\nu$ distributions of the $N_C=1$ subset and of the  $N_C>3$ subset are not discrepant. 
The $p-$values of the KS tests for all comparisons are quoted in Table~\ref{tab:pvalues}, along with the median $\chi^2_{\nu}$ for the fits of each subsample.
Despite the sparse nature of the data (measurements in only $\sim 3.5$ different filters on average per target), the sample is sufficiently large to allow for the statistical distinction of LOSs with two or three clouds from those with only a single cloud.
The deterioration of the Serkowski fits in the multicloud LOSs is mainly driven by the behavior of the EVPA with wavelength, and is made clear thanks to our $q-u$ space fits. In contrast, if we perform the fits only on $P-\lambda$, it is not possible to discriminate between LOSs with and without multiple clouds with current data.
When the structure in the LOS is too complicated ($N_C>3$), there is no statistical difference in the polarization profiles compared to the single cloud case.
This is not surprising, as with an increasing number of clouds the variations induced by individual screens tend to average out.
Even deviations in the EVPA, which should be more prominent, would become smeared and difficult to detect, especially with limited data points across the spectrum.

\begin{table}[!htb]
\caption{$p$ values of the KS test between the subsample of targets with $N_C=1$ and other subsamples (Sect.~\ref{subsec:statistics}).}
\centering
\begin{tabular}{c c c}
Subsamples &  $p$-value & Median $\chi^2_{\nu}$\\
\hline
$N_C=1$ &  1 & 1.20
 \\
$N_C=2$ &  1.5 $\cdot 10^{-3}$ & 2.59 \\
$N_C=3$ &  2.1 $\cdot 10^{-2}$ & 2.17 \\
$N_C=2$ or 3  & $1.8 \cdot 10^{-3}$ & 2.51 \\
$N_C>1$ & 2.1 $\cdot 10^{-2}$ & 1.97 \\
$N_C>3$ &  $1.5 \cdot 10^{-1}$ & 0.86 \\
\end{tabular}
\label{tab:pvalues}
\end{table}

\subsection{Two revealing examples}\label{subsec:examples}

To demonstrate the impact of the LOS variations in $\lambda_{max}$  and EVPA on the Serkowski relation, we employed archival spectropolarimetric data \citep{Bagnulo2017}.
We identified two revealing cases, where the multiwavelength polarization data show clear evidence of the LOS effects.
The examples discussed here are not part of our statistical sample, as their distances are higher than the limitations imposed in Sect.~\ref{subsec:data}.

\begin{figure*}
    \centering
    \includegraphics[width=1.\linewidth]{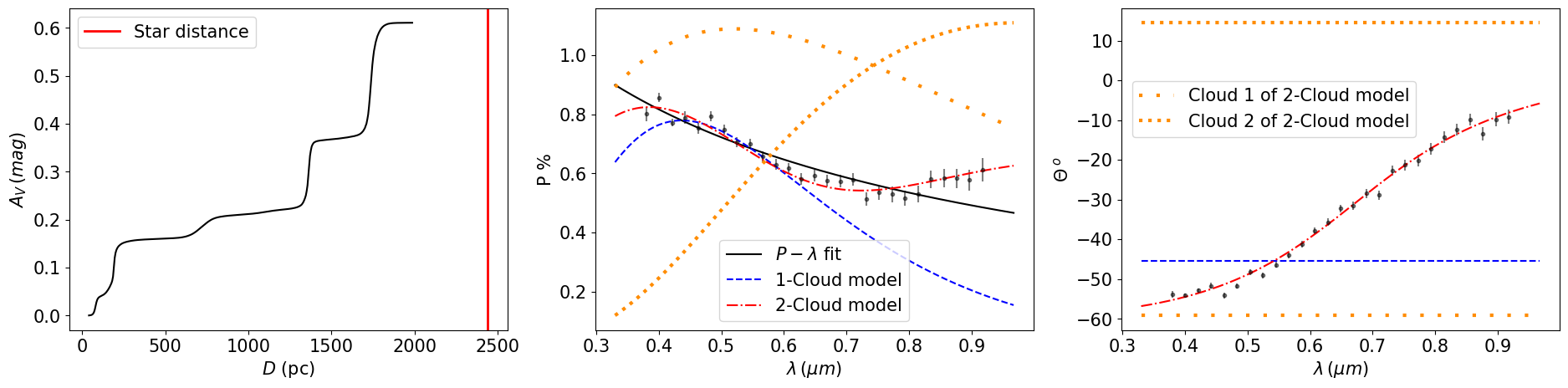}
    \caption{Example of a case in which both the $P-\lambda$ and the $\theta-\lambda$ profiles deviate from Serkowski expectations. We were able to fit both profiles by introducing a 2-Cloud model. Data from \cite{Bagnulo2017} for HD93222. Left: Extinction vs. distance profile taken from the 2kpc version of the map by \cite{Edenhofer2024}. The vertical red line corresponds to the distance of the source. Middle: Data in the $P-\lambda$ space, together with three fit models. Solid black line: Fit using Eq.~\ref{eq:Serkowski} in $P-\lambda$ space only. Dashed blue line: Best-fit 1-Cloud model in the $q-u$ space (Eq.~\ref{eq:quhat}). Dash-dotted red line: Best-fit 2-Cloud model in the $q-u$ space (Eq.~\ref{eq:quhat}). Dotted orange lines correspond to the two individual components of the 2-Cloud model. Right: Data in $\theta-\lambda$ space together with the three models and the individual components of the 2-Cloud model. Parameters of the fits are quoted in Table~\ref{tab:HD163181_params}.}
    \label{fig:HD93222}
\end{figure*}

The LOS toward the main sequence star HD93222 with spectral type O7 represents a prominent example of what the $P-\lambda$, and $\theta - \lambda$ profiles look like when 3D effects are significant.
First, the polarization as a function of wavelength deviates from the typical Serkowski curve (Fig.~\ref{fig:HD93222} middle panel).
Fitting the $P-\lambda$ data with the Serkowski relation requires unphysical values for both $\lambda_{max}$, and $K$ (Table~\ref{tab:HD163181_params}), which still yield a merely adequate fit.
In addition, $\theta$ varies smoothly with wavelength, $\lambda$, which is unexpected when the polarization is dominated by a single cloud with a well-defined mean magnetic field. Thus, both $P$ and EVPA show an atypical behavior with $\lambda$ for this star.
Using 3D dust extinction maps \citep{Edenhofer2024}, we verified the existence of multiple clouds along the LOS toward the HD93222 star.
The clouds are identified as abrupt increments in extinction as a function of distance (left panel in Fig.~\ref{fig:HD93222}).
The average extinction profile shows three prominent clouds located at around 0.25, 1.35, and 1.8 kpc.
The target star, whose estimated distance is 2.44 kpc \citep{BailerJones2021}, lies beyond these clouds, and hence its polarization signal carries information from all of them.
We fit the data with models with different number of clouds.
Although the extinction profile exhibits three distinct steps, the data can be described equally well with both a two- and a three-cloud model. 
This may indicate that two out of three clouds have similar properties, and their constructive profile resembles that of one cloud (such as the case of the middle left panel of Fig.~\ref{fig:toymodels}). 
We proceeded with the simplest case of a two-cloud model.

\begin{table*}[!htb]
    \caption{Fit parameters for the stars HD163181 and HD93222 in Sect.~\ref{subsec:examples}.}
    \centering
    \begin{tabular}{c c c c c c c}
    & $P_{\max}$ \% & $\lambda_{\max}$ ($\mu m$) & $K$ & $\theta$ ($^\circ$) & $K/\lambda_{max}$ ($\mu m ^{-1}$) & $\chi^2_\nu$\\
    \hline \\
    \multicolumn{6}{c}{HD93222} \\
    \hline
    $p-\lambda$ only (This work) & 1.42 & 0.048 & 0.12 & -- & 2.5 & 3.33\\
    1-Cloud $q-u$ Model & 0.78 & 0.437 & 2.55 & $-46.80$ & 5.84 & 91.49\\
    2-Cloud $q-u$ Model & -- & -- & -- & -- & -- & 2.21\\
    2-Cloud Model - Cloud 1 & 1.09 & 0.521 & 0.96 & $-59.15$ & 1.84 & --\\
    2-Cloud Model - Cloud 2 & 1.11 & 0.975 & 1.90 & 14.49 & 1.95 & --\\
    \hline
    \\    
    \multicolumn{6}{c}{HD163181} \\
    \hline
    $p-\lambda$ only \citep{Bagnulo2017} & 1.45 & 0.458 & 0.50 & -- & 1.09 & 1.68 \\
    1-Cloud $q-u$ Model & 1.45 & 0.496 & 0.55 & $-4.15$ & 1.11 & 20.42\\
    2-Cloud $q-u$ Model & -- & -- & -- & -- & -- & 2.46\\
    2-Cloud Model - Cloud 1 & 1.39 & 0.463 & 0.71 & $-19.57$ & 1.53 & --\\
    2-Cloud Model - Cloud 2 & 0.77 & 0.731 & 1.00 & 29.30 & 1.37 & -- \\
    \hline   
    \end{tabular}
    \label{tab:HD163181_params}
\end{table*}

\begin{figure*}
    \centering
    \includegraphics[width=1.\linewidth]{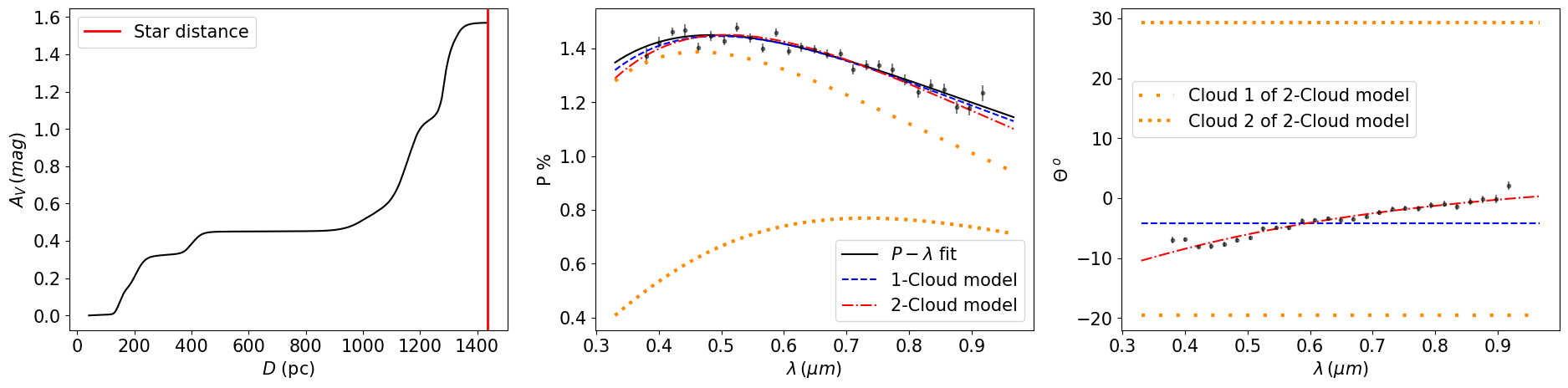}
    \caption{Example of a case in which the $P-\lambda$ profile follows the Serkowski relation, but the EVPA does not. By employing a model with two components in the LOS, it is possible to fit both $P$ and $\theta$. Data from \cite{Bagnulo2017} for HD163181. Panels and lines similar to Fig.~\ref{fig:HD93222}. Parameters of the fits are quoted in Table~\ref{tab:HD163181_params}.}
    \label{fig:HD163181}
\end{figure*}

As a second example, we show results for the star HD163181, which is an intermediate-size luminous supergiant of spectral class O9.5 (Fig.~\ref{fig:HD163181}).
In this case, the observed degree of polarization follows the Serkowski formula, but the EVPA varies with wavelength.
The 3D extinction profile has two significant steps, indicating the existence of more than one dominant cloud along this LOS.
Assuming the standard Serkowski curve with a single dominant cloud can yield good fits to the data in the $P-\lambda$ space \citep{Bagnulo2017}, but it fails to explain the observed variability of the EVPA; we verified this by fitting a one cloud model in the $q-u$ plane (dashed blue line in Fig.~\ref{fig:HD163181}). 
However, when we considered two clouds along this LOS, as the 3D extinction map suggests, we were able to capture the observed variability of both $P$ and $\theta$ with wavelength (dash-dotted red line in Fig.~\ref{fig:HD163181}, middle panel).

\subsection{The parameters of the Serkowski formula}\label{subsec:parameters}

\begin{figure}
    \centering
    \includegraphics[width=0.99\linewidth]{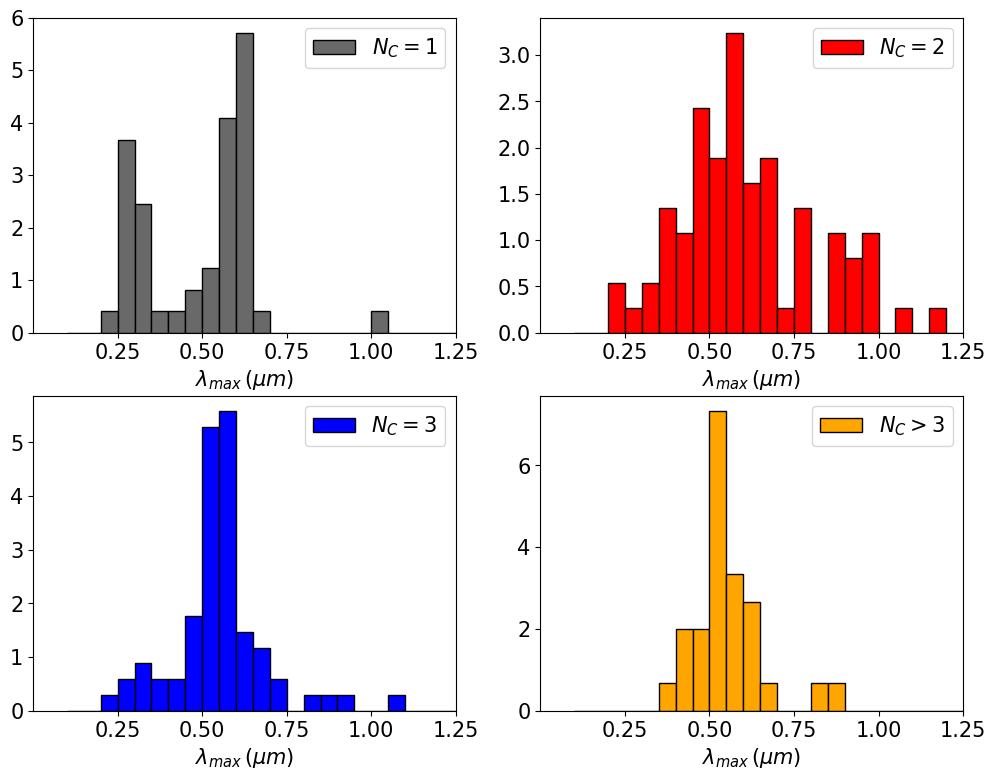}
    \caption{Normalized distributions of $\lambda_{max}$ for subsamples with different number of clouds in the LOS. See Sect.~\ref{subsec:parameters} for details.}
    \label{fig:lmax_per_cloud}
\end{figure}

\begin{figure}
    \centering
    \includegraphics[width=0.99\linewidth]{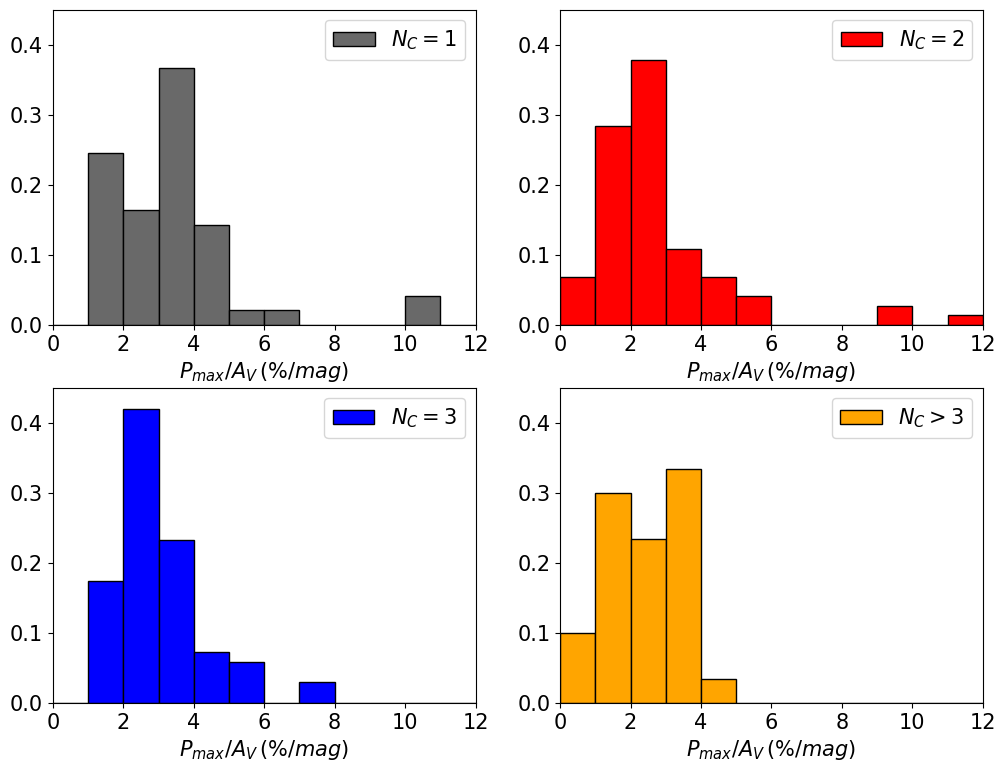}
    \caption{Normalized distributions of $P_{max}$ for subsamples with different number of clouds in the LOS. See Sect.~\ref{subsec:parameters} for details. The following bins are omitted for clarity. For $N_C=2$: bin at $P_{max}/A_V\simeq 23$ with a height of 0.0135. For $N_C=3$: bin at $P_{max}/A_V\simeq 13$ with a height of 0.0145.}
    \label{fig:pmax_per_cloud}
\end{figure}

Here, we demonstrate how the derived parameters of the Serkowski relation can be incorrect, when erroneously assuming a single cloud in the LOS.
We explored how 3D effects affect the distributions of $\lambda_{max}$, and $P_{max}$ that one obtains through the classical fitting of the Serkowski relation in the $P - \lambda$ space; $K$ linearly correlates with $\lambda_{max}$ \citep{Wilking1980}, and thus can be omitted from this analysis.
We fit the data only in $P-\lambda$ space (Eq.~\ref{eq:Serkowski}) by minimizing 
\begin{equation}
  \frac{1}{N_\lambda} \cdot \sum \left( \frac{P_i-\hat{P_i}}{\sigma_{p_i}} \right)^2 ,
\end{equation}
assuming a one-cloud model, and imposing the same priors as discussed in Sect.~\ref{subsec:fitmethod}.
For these fits, we used the debiased $P$ values.
Since the effect of the 3D dust structure is more prominent in $\theta-\lambda$ space, fits in $P-\lambda$ were comparably good for all four subsamples.
We then explored the normalized distributions of the derived parameters for all subsamples (Fig.~\ref{fig:lmax_per_cloud} for $\lambda_{max}$, Fig.~\ref{fig:pmax_per_cloud} for $P_{max}/A_V$).

Past studies of the Serkowski parameters show that the distribution of  $\lambda_{max}$ is singly peaked, with an average value close to 0.5 $\mu$m \citep{martin_whittet_1990}. In contrast, we find that for $N_C=1$ (single-cloud sightlines) the intrinsic distribution of $\lambda_{max}$ is bimodal, with peaks at approximately 0.28 and 0.62 $\mu$m. On the other hand, all distributions with $N_C>1$ (multiple-cloud sightlines) are unimodal and peak at around 0.55 $\mu$m, which is close to the value that is considered to be the Galactic average. By performing MCMC simulations, we found that the unimodality in the $\lambda_{max}$ distribution, which is observed for $N_C > 1$, can be reproduced with multiple-cloud models with values of $\lambda_{max}$ randomly drawn from the $N_C=1$ distribution.
The intrinsic Galactic distribution of $\lambda_{max}$ may thus be bimodal, and the currently accepted value for the Galactic average of $\lambda_{max}$ ($\sim 0.5 \mu$m) is likely contaminated by LOS effects.
Notably, the larger spread of the distribution of $\lambda_{max}$ for the $N_C=2$ case is not surprising.
\cite{Clarke1984} theorized that in the case of two clouds in the LOS, the acquired $\lambda_{max}$ from the Serkowski fit is different from the average $\lambda_{max}$ of the two clouds when the two clouds have different parameters.
Furthermore, it is worth considering that most areas containing stars behind a single cloud also contain stars behind multiple clouds. We calculated the average $\lambda_{max}$ for targets in distinct areas on the sky (Fig.~\ref{fig:skypositions}).
In most cases, the average $\lambda_{max}$ differs significantly between subsets (see Table~\ref{tab:lmax}) located in the same area. This suggests that using the $\lambda_{max}$ of a star from an area with $N_C>1$ as a characteristic value for the LOS dust properties could be misleading.


\begin{table*}[!htb]
\caption{Mean $\lambda_{max}$ values in $\mu m$ and their error on the mean for different regions in the sky and different subsamples.}
\centering
\begin{tabular}{c c c c c}
Region & $N_C=1$ & $N_C=2$ & $N_C=3$ & $N_C>3$ \\
\hline
A & $0.28 \pm 0.02$ & $0.56 \pm 0.06$ &      &      \\
B & $0.34 \pm 0.02$ &      & $0.63 \pm N/A$ &      \\
C & $0.83 \pm 0.14$ & $0.60 \pm 0.04$ &      &      \\
D & $0.54 \pm 0.04$ & $0.28 \pm N/A$ &      &      \\
E & $0.53 \pm N/A$ & $0.54 \pm 0.02$ & $0.55 \pm 0.02$ & $0.55 \pm 0.01$ \\
F & $0.52 \pm N/A$ & $0.52 \pm 0.02$ & $0.52 \pm 0.01$ &       \\
G & $0.56 \pm N/A$ &      &      & $0.49 \pm 0.02$ \\
\hline
\end{tabular}
\tablefoot{Regions with N/A in the error of the mean are cases in which there is only one star in the specific region for the subsample.}
\label{tab:lmax}
\end{table*}

The distribution of $P_{max}/A_V$ is also affected by the number of clouds along the LOS. We can immediately see this by assuming, without loss of generality, that two clouds lie along a LOS. The polarization induced by the two clouds can either (partially or totally) add up, or (partially or totally) cancel out, depending on the difference between the EVPAs each would impart on its own.
When the EVPA difference is close to 0$^o$, the polarization adds up, while when the difference is close to 90$^o$ it cancels out.
On the other hand, the extinction, $A_V$, induced by each cloud is always additive. Therefore, with more clouds, we would expect to have in general lower $P_{max}/A_V$ than in the $N_C=1$ cases. However, since $P_{max}$ is a derived parameter rather than a direct measurement, the fit can be poor in cases with $N_C > 1$, and $P_{max}$ may not represent any physical truth. In such scenarios, $P_{max}$, and thus $P_{max}/A_V$, could take on arbitrary values.
This is exactly what we observe in the distributions of $P_{max}/A_V$ (Fig.~\ref{fig:pmax_per_cloud}). The $P_{max}/A_V$ distribution for $N_C=1$ is free from LOS effects, and hence it should resemble the “ground truth.” 
This distribution peaks at $P_{max}/A_V \simeq 3-3.5 \% /$ mag, with only a few points with $P_{max}/A_V > 4.5 \% /$ mag. On the other hand, the $N_C=2$ distribution peaks at a lower value of $P_{max}/A_V \simeq 2 \% /$ mag, with its tail extending to higher values of $P_{max}/A_V$ than when $N_C=1$.
The shape of the $N_C=2$ distribution exactly follows the expectations discussed above.
Similarly, for the distributions of more components, $P_{max}/A_V$ consistently has lower values than the case of $N_C=1$.

The polarization-to-extinction ratio has garnered significant interest within the ISM community \citep[e.g.][]{Hiltner1956,Serkowski1975,Panopoulou2019,Planck2020,Angarita2023}. While here we present our findings for $P_{max}/A_V$, discussions typically focus on $P/A_V$ in the context of measured polarization in a specific band, rather than on a fit parameter such as $P_{max}$. Consequently, our results for the distributions with $N_C>1$ cannot be fully integrated into this broader discussion if the fit parameter, $P_{max}$, does not accurately reflect the true value for a given field. On the contrary, the findings for the $N_C=1$ case should be free from bias. Within this subsample, we identify three targets with extreme $P_{max}/A_V$ values: two with $P_{max}/A_V \simeq 10-11\% / \mathrm{mag}$ and one with $P_{max}/A_V \simeq 6\% / \mathrm{mag}$. These values correspond to $P_{max}/E(B-V) \simeq 31-34\% / \mathrm{mag}$ and $P_{max}/E(B-V) \simeq 18.6\% / \mathrm{mag}$, respectively, assuming that $R_V = 3.1$ with $R_V = A_V / E(B-V)$ \citep{Fitzpatrick2004}. These values exceed the highest $P_{max}/E(B-V)$ reported by \cite{Angarita2023} to date. However, the observed maximum $P/E(B-V)$ has consistently been revised upwards over time: from $P/E(B-V) \leq 9\% / \mathrm{mag}$ \citep{Hiltner1956,Serkowski1975}, to $P/E(B-V) \leq 13\% / \mathrm{mag}$ \citep{Panopoulou2019,Planck2020}, and most recently to $P/E(B-V) \leq 16\% / \mathrm{mag}$ \citep{Angarita2023}. While our values may appear extreme, they may well reflect true physical conditions, and further research is warranted.

Overall, both the $\lambda_{max}$ and the $P_{max}/A_V$ are significantly affected by the 3D structure of the ISM, which, if ignored, can lead to erroneous Serkowski best-fit parameters.
Our results suggest that fitting a single Serkowski relation is invalid in LOSs with multiple clouds. There can be cases in which a single Serkowski relation yields good fits, even when 3D effects are important (Sect.~\ref{subsec:examples}). However, in these cases the best-fit parameters of a single Serkowski curve are misleading, and not representative of the dust grain physics.

All the targets used in the statistical sample are presented in Table~\ref{tab:targets_CDS} with their properties. The complete table can be found only in electronic form (Sect.\ref{sec:dataavail}).

\begin{table*}[]
\caption{Targets and their properties.}
    \centering
    \begin{tabular}{|c|c|c|c|c|c|c|c|c|c|c|c|}
    \hline
    \textsc{Gaia} ID & RA & Dec & Distance & $A_V$ & $N_{bands}$ & $N_C$ & $\chi^2_\nu$ & $P_{max}$ & $K$ & $\lambda_{max}$ & $\theta$ \\
    & ($\circ$) & ($\circ$) & (pc) & (mag) & & & & $(\%)$ & & ($\mu m$) & ($\circ$) \\
    \hline
    180979373413769344  &  80.68056  &  33.28809  &  1151  &  0.61  &  4  &  >3  &  0.61  &  1.30  &  0.56  &  0.350  &  -16.7 \\
    \hline
    \end{tabular}
    \label{tab:targets_CDS}
\end{table*}
\tablefoot{The full table is available at the CDS.}

\section{Discussion}

\subsection{Implications for dust modeling}
The findings presented in this study have significant implications for dust modeling. Dust models and grain alignment theories must satisfy the constraints obtained from Serkowski fits \citep{martin_whittet_1990,andersson_2015.alignment.review,draine_2024}. However, our results indicate that the Serkowski parameters derived from fitting the observed polarization data in complex interstellar environments may not accurately represent the underlying dust properties, leading to erroneous conclusions about grain size distributions and composition.
The results from Sect.~\ref{subsec:parameters} suggest that the intrinsic distribution of $\lambda_{max}$ in the clouds of our Galaxy may actually be bimodal (Fig.~\ref{fig:lmax_per_cloud}). This finding contradicts past studies \citep{martin_whittet_1990}, which likely stem from (incorrectly) fitting the Serkowski formula in regions with multiple clouds. Although, our employed sample is the largest to date, the statistics are still limited, and more data are required to further explore the intrinsic distribution of $\lambda_{max}$ in our Galaxy. 

We propose the following two options for constraining the populations of dust in different ISM environments: 1) fit the Serkowski formula only to targets behind a single cloud, or 2) decompose the cumulative polarization signal into its individual components for targets behind multiple clouds to derive the characteristics of each.
Moreover, different regions of the Galaxy (e.g., the disk vs. the halo, or varying distances from the Galactic center) may exhibit distinct dust physics that we are currently missing due to the misapplication of the Serkowski relation. By studying the polarization of targets behind multiple clouds and decomposing the signal to derive individual parameters, it may become possible to uncover dust clouds with diverse and potentially novel properties in different parts of the Galaxy.

\subsection{Implications for the identification of intrinsically polarized stars}
It is customary to interpret deviations from the Serkowski relation in the $P(\lambda)$ behavior of point sources as indicative of intrinsic polarization mechanisms, such as the presence of circumstellar disks \citep[e.g.][]{Topasna2023}.
However, the wavelength dependence of polarization in such cases is known to lack a typical pattern \citep{Bastien2015}.
In contrast, we have demonstrated that fitting simple two-cloud models can account for deviations from the Serkowski formula quite effectively
(Sect.~\ref{subsec:examples}).
This suggests that deviations from the Serkowski curve alone are insufficient to indicate the presence of an intrinsic polarization component.
In order to confirm that a star not adhering to the Serkowski formula is intrinsically polarized, other evidence, such as the detection of variability in the polarization or a difference in polarization between the star and its neighbors, is required.

\subsection{The possibility of 3D tomography with the Serkowski relation}\label{subsec:tomography}

Fitting multiple-cloud models to multiwavelength (optical to near-infrared) polarization data paves the way for constraining the LOS variations in the plane-of-the-sky magnetic field morphology (tomography).
If the number of clouds in the LOS is known, for instance from 3D extinction maps \citep[e.g.,][]{Edenhofer2024, green_2019.3D.extinction.map}, it is possible to extract the parameters of individual clouds using the techniques discussed in Sect.~\ref{subsec:multiple_component_fits}.
However, the following considerations should be kept in mind.
It is possible that two or more clouds in the LOS have similar parameters.
In such cases, the polarization data may be adequately fit with fewer clouds than what the 3D extinction maps suggest. For example, if two clouds along a LOS have comparable $\lambda_{max}$ and $\theta$, their polarization contributions will combine, resulting in a profile resembling that of a single cloud but with an increased degree of polarization.
This might explain our target star HD93222 (Sect.~\ref{subsec:examples}), where the extinction profile displays at least three distinct steps, yet the data are sufficiently well fit by a 2-Cloud model.
Another possibility is that the third cloud has a low abundance of aligned grains, because of, for example, a magnetic field with a direction along the LOS.

Another strategy one could follow to identify cases of clouds with comparable properties is to obtain multiband or, ideally, spectropolarimetric measurements of multiple stars at different distances along the same LOS.
The polarization signal would vary with distance, as each star would only be influenced by the foreground dust column.
Thus, stars lying behind the same clouds would show similar polarization trends, while distinct changes would only occur for stars behind different clouds.
Following such methods could complement the BISP-1 algorithm \citep{pelgrims_2023}, which employs single-band optical polarization data with parallaxes from Gaia.
This algorithm is particularly relevant for performing ISM tomography along many LOSs \citep{Pelgrims2024}, especially in the context of optopolarimetric surveys, such as the \textsc{Pasiphae} survey \citep{tassis_2018.pasiphae}.
While the method described by \cite{pelgrims_2023} can determine the number of clouds along a LOS and their distances, it does not fully capture the dust physics within each cloud. Multiband polarimetric measurements of a few stars at varying distances along the LOS can provide insights into the dust physics within each cloud.
Thus, multiwavelength polarization data promises to provide an independent method for performing ISM magnetic field tomography, which could usher in the era of high-precision 3D magnetized ISM cartography.

\section{Summary and conclusions}

In this paper, we showcase the effect of the 3D structure of the ISM on the well-known and widely used Serkowski relation.
Our findings are summarized as follows.
\begin{enumerate}
    \item We have revisited theoretically how the Serkowski formula could become invalid in LOSs with many dust components. We have shown that the variations in both the degree of polarization and EVPA with wavelength can be severely affected in some stars. In other stars, the polarization profile may misleadingly appear to follow the Serkowski relation, despite the contribution from multiple dust clouds. In such cases, the imprint of the 3D structure can mainly be observed in the $\theta-\lambda$ profile.

    \item As a test case, we selected two different targets that appear to be behind multiple dust clouds, based on their extinction profiles, and for which archival spectropolarimetric data exist.
    We demonstrated that the single-cloud model, implicitly assumed for Serkowski fits, fails in these cases, while we successfully fit two-cloud models.

    \item We propose the $q-u$ fitting technique as the most appropriate approach to fit the Serkowski relation. This method allows one to trace both $P$ and $\theta$ as functions of wavelength simultaneously, enabling fitting for more than one cloud along the LOS.

    \item We used a sample of 223 stars behind different numbers of clouds with polarimetric measurements in three or more bands, to fit and evaluate statistically the performance of the Serkowski formula. We  found that the Serkowski formula is not a good model for LOSs with two or more clouds, while it fits the data well in the cases of one or more than three clouds in the LOS. The latter is an outcome of averaging. 

    \item Our analysis of the fit parameters of the 223 stars suggests that the intrinsic distribution of the $\lambda_{max}$ parameter in our Galaxy may be bimodal, something that was veiled until now, due to treating sightlines with many clouds as a single dust component.
    
    \item A poor Serkowski fit does not by itself constitute conclusive evidence of intrinsic polarization in a source.
    
    \item Fitting the Serkowski formula for stars behind multiple dust clouds leads to an incorrect estimation of the fit parameters, and thus of the underlying dust physics.

    \item Our results hint at the possibility of performing magnetic tomography using multiband polarization data in combination with \textsc{Gaia} distances. 
\end{enumerate}

In conclusion, the Serkowski relation should be cautiously applied when the ISM 3D structure is complex. Finally, our findings underscore the critical role of spectropolarimetry in driving future advancements in ISM research. While wide-field polarimetry offers valuable insights, its limitations become evident when considering the physics of dust. Therefore, we strongly advocate for the widespread adoption of spectropolarimetry within the ISM community for future research.

\section*{Data availability} \label{sec:dataavail}
Table~\ref{tab:targets_CDS} is only available in electronic form at the CDS via anonymous ftp to \url{cdsarc.u-strasbg.fr (130.79.128.5)} or via \url{http://cdsweb.u-strasbg.fr/cgi-bin/qcat?J/A+A/}

\begin{acknowledgements}
The authors thank V. Pavlidou and G. Panopoulou for their constructive feedback. The authors also thank B. Hensley and A. Readhead for very fruitful discussions.
NM and KT were supported by the European Research Council (ERC) under grant agreements No. 7712821. This work was supported by NSF grant AST-2109127.
N.M. was funded by the European Union ERC-2022-STG - BOOTES - 101076343. Views and opinions expressed are however those of the author(s) only and do not necessarily reflect those of the European Union or the European Research Council Executive Agency. Neither the European Union nor the granting authority can be held responsible for them.
Support for this work was provided by NASA through the NASA Hubble Fellowship grant \#~HST-HF2-51566.001 awarded by the Space Telescope Science Institute, which is operated by the Association of Universities for Research in Astronomy, Inc., for NASA, under contract NAS5-26555.
\end{acknowledgements}

\bibliographystyle{aa}
\bibliography{bibliography}

\end{document}